\documentclass{article}
\usepackage{frascatiphys}
\usepackage{graphicx}
\newcommand{\apjl}{Astrophys. J. Lett.} 
\newcommand{\apj}{Astrophys. J.} 
 
\newcommand{\app}{Astropart. Phys. }

\newcommand{\prd}{Physics Rev. D }

\newcommand{\etal}{et al.}

\begin{document}
\title{ 
MEASUREMENT OF THE COSMIC RAY ENERGY SPECTRUM WITH ARGO-YBJ
}
\author{
Giuseppe Di Sciascio on behalf of the ARGO-YBJ Collaboration   \\
{\em INFN, Sezione di Roma Tor Vergata, Italy} \\
}
\maketitle
\baselineskip=11.6pt
\begin{abstract}
The ARGO-YBJ detector, located at high altitude in the Cosmic Ray Observatory of Yangbajing in Tibet (4300 m asl, about 600 g/cm$^2$ of atmospheric depth) provides the opportunity to study, with unprecedented resolution, the cosmic ray physics in the primary energy region between 10$^{12}$ and 10$^{16}$ eV. The preliminary results of the measurement of all-particle and light-component (i.e. protons and helium) energy spectra between approximately 5 TeV and 5 PeV are reported and discussed. The study of such energy region is particularly interesting because not only it allows a better understanding of the so called 'knee' of the energy spectrum and of its origin, but also provides a powerful cross-check among very different experimental techniques. The comparison between direct measurements by balloons/satellites and the results by surface detectors, implying the knowledge of shower development in the atmosphere, also allows to test the hadronic interaction models currently used for understanding particle and cosmic ray physics up the highest energies.
\end{abstract}
\baselineskip=14pt

\section{Introduction}

There is a general consensus that Galactic cosmic rays (hereafter CRs) up to the ``knee'' ($\sim$3--4$\cdot$10$^{15}$ eV) originate in Supernova Remnants (SNRs) accelerated by the first order Fermi mechanism in shock waves. The theoretical modelling of this mechanism can reproduce the measured spectra and composition of CRs. 
Recent measurements carried out by the balloon-borne CREAM experiment \cite{cream1,cream2} show that the proton and helium spectra from 2.5 to 250 TeV are both flatter compared to the lower energy measurements. In particular, the proton spectrum in this energy range is found harder than the value obtained by fitting many previous direct measurements \cite{horandel}. 
The evolution of the proton and helium spectra and their subtle differences can be an indication of the contribution of different populations of CR sources operating in environments with different chemical compositions \cite{blasi11}.

In the knee region the measurements of the CR primary spectrum are carried out only by EAS arrays and the current experimental results are conflicting. In the standard picture the "mass of the knee" is light being due to the steepening of the p and He spectra \cite{kascade}. 
However, different experiments attribute the "mass of the knee" to higher nuclei. 
A hybrid measurement carried out exploiting the Cherenkov light yield detected by the EAS-TOP experiment (located at 2000 m a.s.l.) at different core distances in EAS and the high energy underground muons sampled by the MACRO experiment, has been used to infer the helium flux at 80 TeV, resulting twice larger than that obtained by JACEE \cite{jacee,aglietta04}.
The EAS-TOP/MACRO analysis implies a decreasing proton contribution to the primary flux well below the observed knee in the primary spectrum. Such considerations can be described through the ratios of the three components at 250 TeV, that can be expressed as: $J_p$ : $J_{He}$ : $J_{CNO}$ = (0.20$\pm$0.08) : (0.58$\pm$0.19) : (0.22$\pm$0.17) \cite{aglietta04}.
In addition, also the results of the Tibet AS$\gamma$ and the BASJE experiments, located at 4300 m a.s.l and at 5200 m a.s.l. respectively, favour a heavier composition because the proton component is no more dominant at the knee \cite{tibet,basje}.
Indications for a substantial fraction of nuclei heavier than helium at 10$^{15}$ eV have been obtained in  old measurements of delayed hadrons \cite{delhad}, as well as by the CASA-MIA collaboration \cite{casamia}.

The knowledge of the primary proton spectrum is fundamental to understand the cosmic rays acceleration mechanisms and the propagation processes in the Galaxy. A careful measurement of the proton spectrum in the energy region from TeV to 10 PeV is the key component for understanding the origin of the knee.
In addition, precise knowledge of its flux may allow one to calculate the yield of rare secondary CRs as antiprotons and positrons and establish the expected fluxes of the atmospheric neutrinos. 

A measurement of the CR primary energy spectrum (all-particle and light component) in the energy range few TeV -- 10 PeV is under way with the ARGO-YBJ experiment (for a description of the detector and a report of the latest physics results see \cite{taup13}). 
To cover this wide energy range different 'eyes' have been used:
\begin{itemize}
\item \emph{'digital readout'}, based on the strip multiplicity, in the few TeV -- 200 TeV energy range \cite{bartoli12};
\item \emph{'analog readout'}, based on the particle density in the shower core region, in the 100 TeV -- 10 PeV range;
\item \emph{'hybrid measurement'}, carried out by ARGO-YBJ and a wide field of view Cherenkov telescope, in the 100 TeV - PeV region \cite{argo-wfcta}.
\end{itemize}
The results concerning the all-particle and the light component (p+He) spectra obtained with the analog readout are summarized in the following. The results obtained with the 'hybrid measurement' are described in \cite{argo-wfcta,caozhen}.

\section{Measurement of the CR light component (p+He) spectrum}

A measurement of the primary CR light (p+He) component energy spectrum has been carried out in the energy range 5 -- 200 TeV exploiting the \emph{digital read-out} of the ARGO-YBJ experiment, i.e. the picture of the EAS provided by the strip/pad system.
With this analysis for the first time a ground-based measurement of the CR spectrum overlaps data obtained with direct methods for more than one energy decade, thus providing a solid anchorage to calibrate the energy scale of EAS arrays approaching the knee region.

%
\begin{figure}
\centerline{\includegraphics[width=0.75\textwidth,clip]{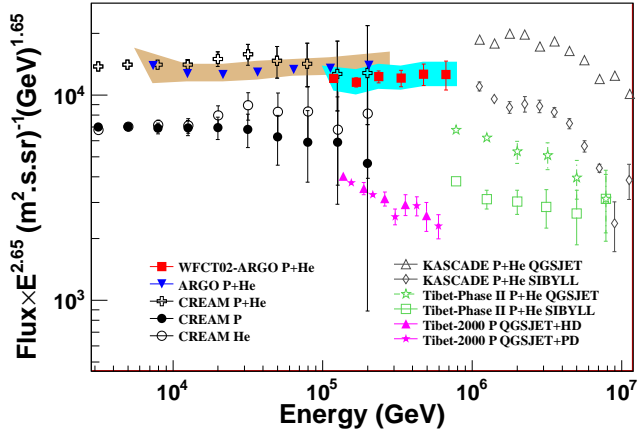} }
\caption{Light component (p+He) energy spectrum of primary CRs measured by ARGO-YBJ compared with other experimental results. The ARGO-YBJ 2012 data refer to the results published in \cite{bartoli12}.
The results obtained by ARGO-YBJ/WFCTA hybrid measurement are shown by the filled red squares \cite{argo-wfcta}.}
\label{fig:light_spectrum}       
\end{figure}
%

The ARGO-YBJ spectrum, reconstructed with an unfolding technique based on the Bayesian approach, agrees remarkably well with the values obtained by adding up the p and He fluxes measured by CREAM both concerning the total intensities and the spectral index. The value of the spectral index of the power-law fit to the ARGO-YBJ data is -2.61$\pm$0.04, which should be compared with $\gamma_p$ = -2.66$\pm$0.02 and $\gamma_{He}$ = -2.58$\pm$0.02 obtained by CREAM \cite{cream2}.
The present analysis does not allow the determination of the individual p and He contribution to the measured flux, but the ARGO-YBJ data clearly exclude the RUNJOB results \cite{rj,runjob}. 
Details can be found in \cite{bartoli12}.

This measurement has been extended to higher energies exploiting an \emph{hybrid measurement} with a prototype of the future Wide Field of view Cherenkov Telescope Array (WFCTA) of the LHAASO project \cite{lhaaso}.
The idea is to combine in a multiparametric analysis two mass-sensitive parameters: the particle density in the shower core measured by the analog readout of ARGO-YBJ and the shape of the Cherenkov footprint measured by WFCTA \cite{argo-wfcta}.
For a detailed description of the technique see \cite{argo-wfcta,caozhen}.

The light component energy spectra measured by ARGO-YBJ up to about 600 TeV with the digital and the hybrid systems are shown in the Fig. \ref{fig:light_spectrum}. 
The hybrid spectrum can be described by a single power-law with a spectral index of -2.63 $\pm$ 0.06 up to about 600 TeV. A systematic uncertainty in the absolute flux of 15\% is shown by the shaded area. The error bars show the statistical errors only.
The absolute flux at 400 TeV is (1.79$\pm$0.16)$\times$10$^{-11}$ GeV$^{-1}$ m$^{-2}$ sr$^{-1}$ s$^{-1}$. 
This result is consistent for what concern spectral index and absolute flux with the measurements carried out by ARGO-YBJ below 200 TeV and by CREAM. The flux difference is about 10\% and can be explained with a difference in the experiments energy scale of $\pm$3.5\% \cite{argo-wfcta}. 

This result is very important to fix the energy scale of the experiment. Below 10 TeV the absolute energy scale of ARGO-YBJ is calibrated at 10\% level exploting the westward displacement of the Moon shadow under the effect of the GMF \cite{argo-moon}. Above this energy the overposition with CREAM allows to compare both energy scales: the agreement is at a few percent level.

\section{All-particle Energy Spectrum in the PeV energy region}

The measurement of the CR energy spectrum up to 10 PeV is under way exploiting the RPC charge readout of the ARGO-YBJ detector which allows to study the structure of the particle density distribution in the shower core region up to particle densities of about 10$^{4}$/m$^2$ \cite{argo-bigpad1,argo-bigpad2}.

The study of the lateral density function (LDF) at ground is expected to provide information on the longitudinal profile of the showers in the atmosphere, that is to estimate their development stage, or \emph{age}, which is related to $X_{max}$, the atmospheric depth at which the cascade reaches its maximum size. This implies the possibility of selecting showers within given intervals of $X_{max}$ or, equivalently, of $X_{dm}$, the distance of the shower maximum from the detector.

The shower development stage in the atmosphere, as observed at a fixed altitude (the detection one), depends on the
energy of the interacting primary. For fixed energy, it depends on the nature of the primary: heavy primaries interact
higher in the atmosphere, thus giving showers which, on average, reach their maximum at a larger distance from 
the detector than a lighter primary of the same energy. 
For this reason, the combined use of the shower energy and age estimations can ensure a sensitivity to the primary mass, thus giving the possibility of selecting a light (p+He) event sample with high efficiency.

%
\begin{figure}[t!]
\begin{minipage}[t]{.47\linewidth}
  \centerline{\includegraphics[width=\textwidth]{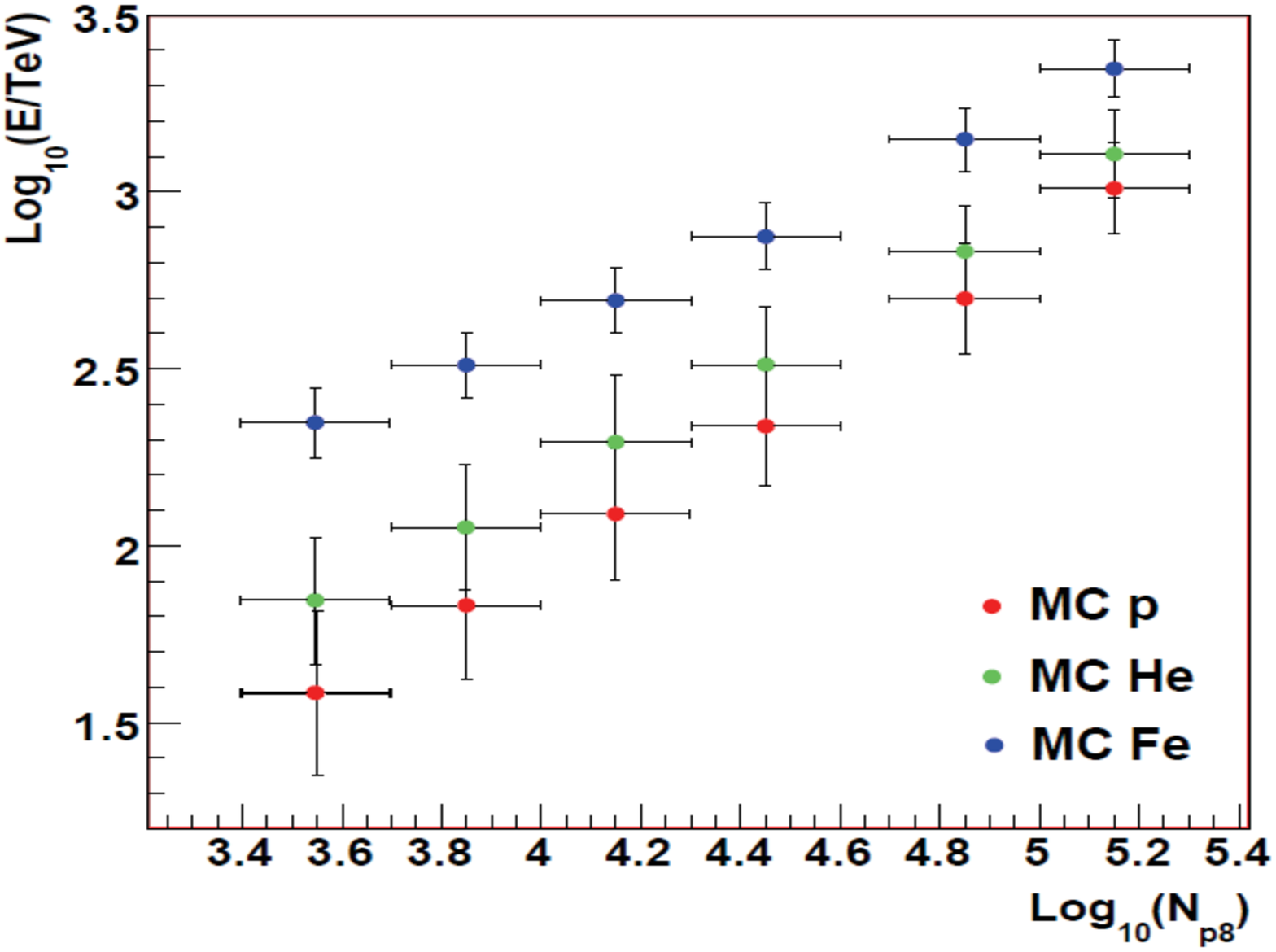} }
\caption[h]{Truncated size N$_{p8}$ as a function of the primary energy for shower induced by different nuclei.}
\label{fig:np8}
\end{minipage}\hfill
\begin{minipage}[t]{.47\linewidth}
  \centerline{\includegraphics[width=\textwidth]{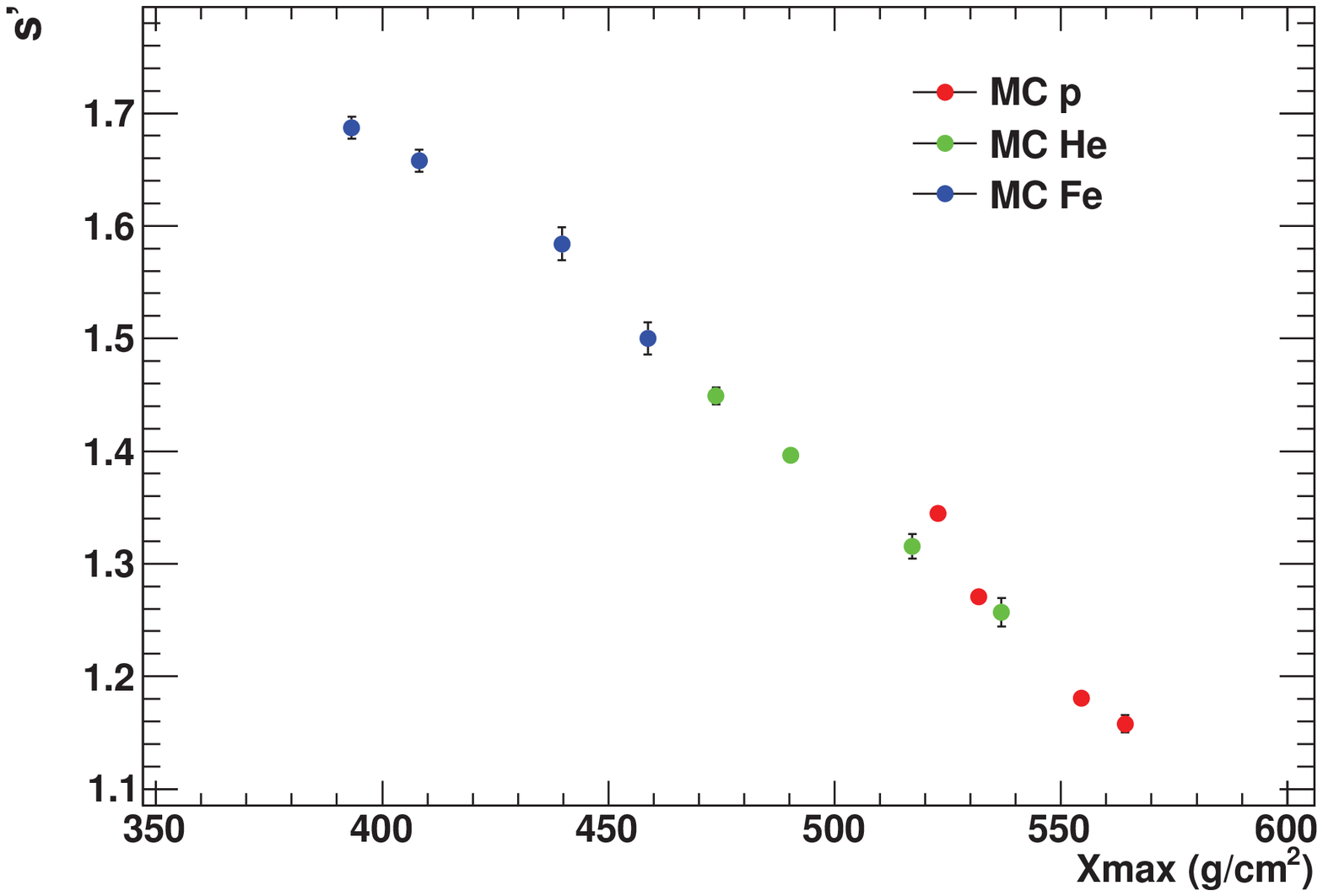} }
\caption{The age parameter $s'$ resulting from the fits of the average LDF of simulated p, He and Fe 
  samples (in each N$_{p8}$ bin) vs the corresponding $X_{max}$ average values.} 
\label{fig:s_Xmax_pHeFe}
\end{minipage}\hfill
\end{figure}
%

Various observables were considered and analyzed in order to find a suitable estimator of the primary CR energy. Among them, according to MC simulations, N$_{p8}$, the number of particles detected within a distance of 8\,m from the shower axis, 
resulted well correlated with energy, not biased by the finite detector size and not much affected by shower to shower fluctuations \cite{icrc779}. Therefore, the analysis is carried out in terms of different N$_{p8}$ intervals to select event samples corresponding to different primary energies. Nevertheless, as shown in Fig. \ref{fig:np8}, this truncated size is a mass-dependent energy estimator parameter.

In order to have a mass-independent parameter we fitted the LDFs of triggered showers (up to about 10\,m from the core) event-by-event, for different N$_{p8}$ intervals and different shower initiating primaries, with a suitable function to get the shape parameter $s'$ (see \cite{icrc781} for details).
From these studies we find that, for a given primary, the $s'$ value decreases when N$_{p8}$ (i.e. the energy) increases, this being due to the observation of younger (deeper) showers at larger energies.
Moreover, for a given range of N$_{p8}$, $s'$ increases going from proton to iron, as a consequence of a larger primary interaction cross section.
Both dependencies are in agreement with the expectations, the slope $s'$ being correlated with the shower age, thus reflecting its development stage.
This outcome has two important implications, since the measurements of $s'$ and N$_{p8}$ can both (i) help constraining the shower age and (ii) give information on the primary particle nature.

Concerning the first point, we show in Fig. \ref{fig:s_Xmax_pHeFe} the $s'$ values as obtained from the fit of the average LDFs, for each simulated primary type and N$_{p8}$ interval, as a function of the corresponding $X_{max}$ average value.
As can be seen, the shape parameter $s'$ depends only on the development stage of the shower, independently from the nature of the primary particle and energy.
That plot expresses an important universality of the LDF of detected EAS particles in terms of the lateral shower age. 
The LDF slope $s'$ is a $X_{max}$ average value estimator mass-independent.
This implies the possibility to select most deeply penetrating showers (and quasi-constant $X_{dm}$ intervals) at different zenith angles, an important point for correlating the exponential angular rate distribution with the interaction length of the initiating particle \cite{aielli09}. Obviously shower-to-shower fluctuations introduce unavoidable systematics, whose effects can be anyway quantified and taken into account.

The second implication is that $s'$ from the LDF fit very close to the shower axis, together with the measurement of the truncated size N$_{p8}$, can give information on the primary particle nature, thus making possible the study of primary mass composition and the selection of a light component data sample.

Assuming an exponential absorption after the shower maximum, we get the size at maximum (N$_{p8}^{max}$) by using N$_{p8}$ and $s'$ measurements for each event: $N_{p8}^{max}\approx\ N_{p8}\cdot exp[(h_0 sec\theta - X_{max}(s'))/\lambda_{abs}]$. A suitable choice of the absorption lenght $\lambda_{abs}$ (=120 g/cm$^2$) allows to get N$_{p8}^{max}$ a parameter correlated with primary energy in an almost linear and mass independent way, providing an energy estimator with a Log(E/TeV) resolution of 0.10--0.15 (getting better with energy). 

%
\begin{figure}
\centerline{\includegraphics[width=0.9\textwidth,clip]{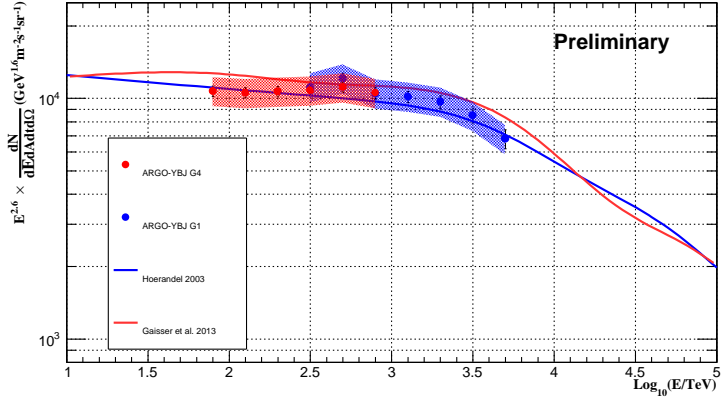} }
\caption{All-particle energy spectrum of primary CRs measured by ARGO-YBJ. Quasi-vertical events ($\theta$ $<$ 15$^{\circ}$)  recorded with two different gain scales (G1 and G4) are plotted. The systematic uncertainty is shown by the shaded area and the statistical one by the error bars. The parametrizations provided by Horandel \cite{horandel} and Gaisser-Stanev-Tilav \cite{gst} are shown for comparison. }
\label{fig:allpart}       
\end{figure}
%

As described in \cite{argo-bigpad1,argo-bigpad2}, with the RPC charge readout we took data with 4 different gain scales to explore the particle density range $\approx$20 -- 10$^4$ particles/m$^2$.
In this preliminary analysis the results obtained with the two intermediate gain scales (so-called G1 and G4) are presented.

Selecting quasi-vertical events ($\theta$ $<$ 15$^{\circ}$) in terms of the truncated size N$_{p8}$ with the described procedure we reconstructed the CR all-particle energy spectrum shown in the Fig. \ref{fig:allpart} in the energy range 100 -- 3000 TeV. 
In the plot a $\pm$15\% systematic uncertainty, due to hadronic interaction models, selection criteria, unfolding algorithms, aperture calculation and energy scale, is shown by the shaded area. The statistical uncertainty is shown by the error bars.
As can be seen from the figure, the two gain scales overlap making us confident about the event selection and the analysis procedure. The ARGO-YBJ all-particle spectrum is in fair agreement with the parametrizations provided by Horandel \cite{horandel} and Gaisser-Stanev-Tilav \cite{gst}, showing evidence of a spectral index change at an energy consistent with the position of the knee. 
 
\section{Observation of the knee in the (p+He) energy spectrum}

%
\begin{figure}
\centerline{\includegraphics[width=0.75\textwidth,clip]{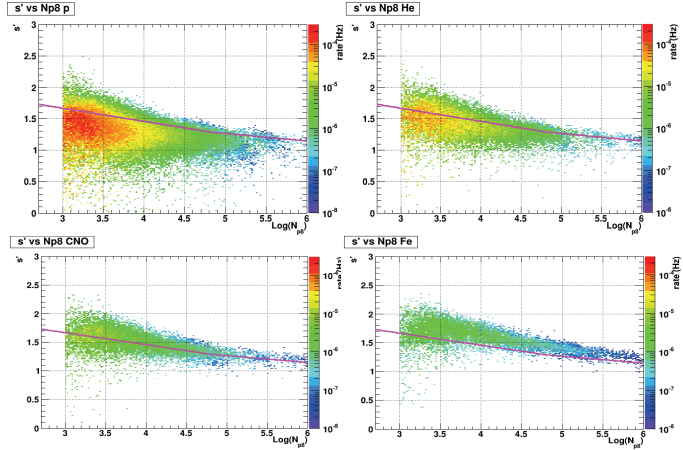} }
\caption{Relation between the LDF shape parameter $s'$ and the truncated size $N_{p8}$ for different nuclei. Showers have been sampled with energy spectra according to Horandel \cite{horandel}. The p+He selection cut is shown by the lines.}
\label{fig:lightsel}       
\end{figure}
%
The measurement of the light component energy spectrum has been extended up to PeVs exploiting three different approaches.
\begin{itemize}
\item[(1)] A selection of events in the $s'$ -- $N_{p8}$ space allowing to get a light (p+He) component sample of showers with a contamination of heavier nuclei of about 15\% (see Fig. \ref{fig:lightsel}). 
\item[(2)] A Bayesian unfolding technique similar to that applied to reconstruct the CR energy spectrum up to 200 TeV. A similar event selection based on the particle density on the central carpet, slightly modified to take into account larger showers recorded with the RPC charge readout, selects a light component event sample with a contamination of heavier nuclei less than 15\%. 
\item[(3)] The ARGO-YBJ/WFCTA hybrid measurement with a different selection procedure which increases the aperture of a factor 2.4 (see \cite{caozhen} for a detailed description of the method and a discussion of the results).
\end{itemize}

\begin{figure}
\centerline{\includegraphics[width=0.9\textwidth,clip]{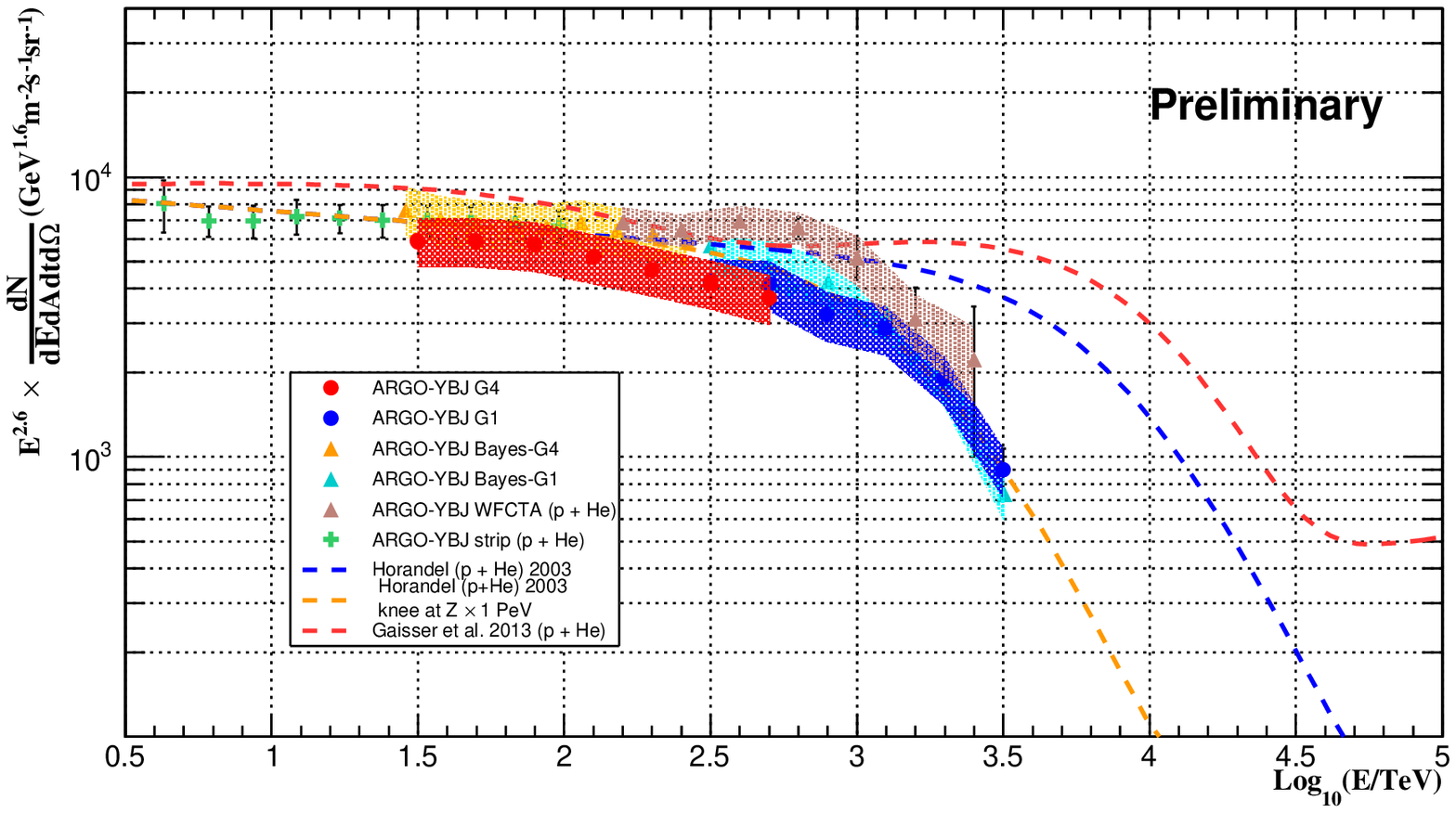} }
\caption{Light (p+He) component energy spectrum of primary CRs measured by ARGO-YBJ with three different analysis. Data recorded with two different gain scales (G1 and G4) are plotted. The systematic uncertainty is shown by the shaded area and the statistical one by the error bars. The parametrizations provided by Horandel \cite{horandel} and Gaisser-Stanev-Tilav \cite{gst} are shown for comparison. A Horandel-like spectrum with a modified knee at Z$\times$1 PeV is also shown.}
\label{fig:phe-knee}       
\centerline{\includegraphics[width=0.9\textwidth,clip]{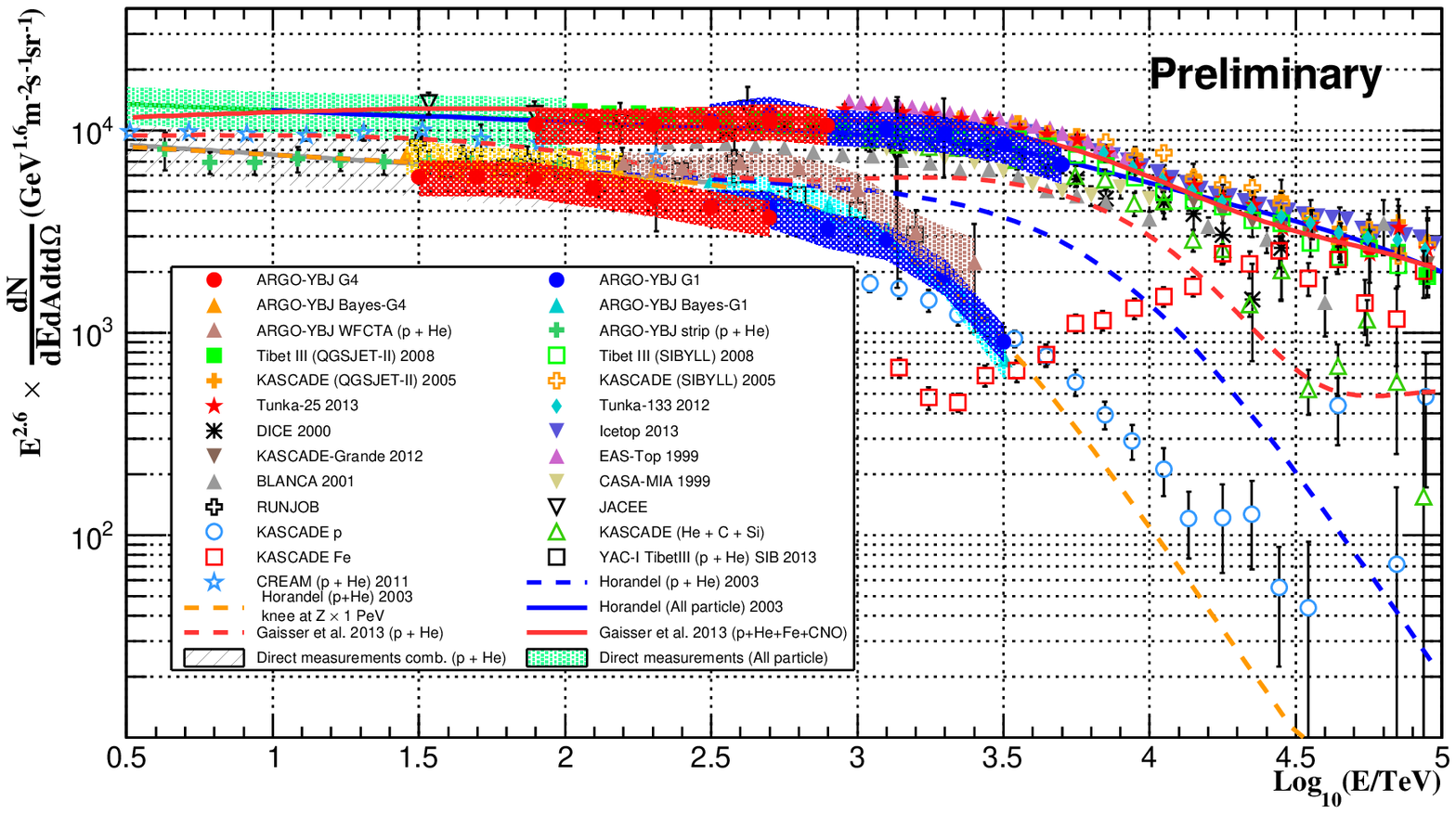} }
\caption{All particle and light (p+He) component energy spectra of primary CR measured by ARGO-YBJ and compared to different experimental results. The parametrizations provided by Horandel \cite{horandel} and Gaisser-Stanev-Tilav \cite{gst} are shown for comparison. A Horandel-like spectrum with a modified knee at Z$\times$1 PeV is also shown.}
\label{fig:all} 
\end{figure}

The energy spectrum of the p+He component measured by ARGO-YBJ with the different methods is summarized in the Fig. \ref{fig:phe-knee}. 
The systematic uncertainty is shown by the shaded area and the statistical one by the error bars.

As can be seen, all three different analysis show evidence of a knee-like structure starting from about 650 TeV. With respect to a single power-law with a spectral index --2.62 the deviation is observed at a level of about 6 s.d. .
The results obtained with the analysis of RPC charge readout data are in fair agreement. These results agree with the ARGO-YBJ/WFCTA hybrid measurement within systematic uncertainty. For comparison, the parametrizations of the light component provided by Horandel \cite{horandel} and Gaisser-Stanev-Tilav \cite{gst} are shown by the blue and red dashed lines, respectively. A Horandel-like spectrum with a modified knee at Z$\times$1 PeV is also shown. 

The all particle and the light component energy spectra measured by ARGO-YBJ are compared to a compilation of different experimental results in the Fig. \ref{fig:all}.

%
%

\section{Conclusions}

The CR energy spectrum has been studied by the ARGO-YBJ experiment in a wide energy range (TeVs $\to$ PeVs) exploiting different approaches. The Moon shadow technique and the overposition with the CREAM data allow to fix the absolute energy scale of ARGO-YBJ up to 4\% level. The all-particle spectrum measured in the energy range 100 -- 3000 TeV is in good agreement with well-known parametrizations, making us confident about the selection and reconstruction of the analog data.
The light component (p+He) has been reconstructed with high resolution up to about 5 PeV. The ARGO-YBJ results show a clear indication of a knee-like structure starting at about 650 TeV. 
Improvements of event selection with the full statistics is under way to extend the measurement up to 10 PeV.
Preliminary results obtained with the last analog gain scale (able to extend the energy range of the charge readout by a factor of 2 at least) are consistent with the results presented in this paper.

\end{document}